\documentclass[12pt,a4paper]{article}
\usepackage{amsmath} 
\usepackage{mathtools}
\usepackage{amsfonts}
\usepackage{amssymb}
\usepackage{graphicx} 
\usepackage {anysize} 
\usepackage{natbib}
\title{\textbf{An analytical effective excess charge density model to predict the streaming potential generated by unsaturated flow}}

\author{Mariangeles Soldi$^{1,*}$, Damien Jougnot$^{2}$, Luis Guarracino$^{1,3}$}
		
\begin{document}
\marginsize{2.5cm}{2cm}{2cm}{2cm} 

\date{}
\maketitle

(1) Consejo Nacional de Investigaciones Cient\'ificas y T\'ecnicas, Facultad de Ciencias Astron\'omicas y Geof\'isicas, Universidad Nacional de La Plata, La Plata, Argentina      
              
(2) Sorbonne Universit\'e,  CNRS, EPHE, UMR 7619 Metis, F-75005, Paris, France 

(3) Facultad de Ciencias Naturales y Museo, Universidad Nacional de La Plata, La Plata, Argentina 

(*)Corresponding author: msoldi@fcaglp.unlp.edu.ar

\vskip 10cm
\noindent
\textit{This paper has been accepted for publication by Geophysical Journal International:}
\\

\noindent
Soldi, M., Jougnot, D., $\&$ Guarracino, L. (2019). An analytical effective excess charge density model to predict the streaming potential generated by unsaturated flow. \textit{Geophysical Journal International}, 216(1), 380-394. DOI: 10.1093/gji/ggy391.

\newpage

\begin{abstract}
The self-potential (SP) method is a passive geophysical method that relies on the measurement of naturally occurring electrical field. One of the contributions to the SP signal is the streaming potential, which is of particular interest in hydrogeophysics as it is directly related to both the water flow and porous medium properties. The streaming current is generated by the relative displacement of an excess of electrical charges located in the electrical double layer surrounding the minerals of the porous media. In this study, we develop a physically based analytical model to estimate the effective excess charge density dragged by the water flow under partially saturated conditions. The proposed model is based on the assumption that the porous media can be represented by a bundle of tortuous capillary tubes with a fractal pore size distribution. The excess charge that is effectively dragged by the water flow is estimated using a flux averaging approach. Under these hypotheses, this new model describes the effective excess charge density as a function of saturation and relative permeability while also depending on the chemical and interface properties, and on petrophysical parameters of the media. The expression of the model has an analytical single closed-form which is consistent with a previous model developed from a different approach. The performance of the proposed model is then tested against previous models and different sets of laboratory and field data from the literature. The predictions of the proposed model fits fairly well the experimental data and shows improvements to estimate the magnitude of the effective excess charge density over the previous models. A relationship between the effective excess charge density and permeability can also be derived from the proposed model, representing a generalization to unsaturated conditions of a widely used empirical relationship. This new model proposes a simple and efficient way to model the streaming current generation for partially saturated porous media.
\\

\end{abstract}

\newpage

\section{Introduction}
\label{sec:intro}
The self-potential (SP) method has gained a strong interest in reservoir and environmental studies due its sensitivity to water flow. Among many other applications of SP, one can mention its use to monitor water flow in the subsurface \citep[e.g.,][]{doussan2002variations,darnet2004modelling,jardani2007tomography,linde2011self,jougnot2015monitoring}, geothermal systems \citep[e.g.,][]{corwin1979self,revil1999streaminggeothermal}, oil and gas reservoirs \citep[e.g.,][]{saunders2006new}, and CO$_2$ sequestration \citep[e.g.,][]{moore2004streaming,busing2017numerical}. Although SP signals are relatively easy to measure, the recorded SP signals are a superposition of different contributions related to redox, diffusion and electrokinetic processes \citep[for more details, see][]{revil2013self}. 

In the present work, we focus on the electrokinetic (EK) contribution to the total SP signal which is directly linked to water flow in porous media and is often referred to as streaming potential. Mineral surfaces are generally electrically charged, creating an electrical double layer (EDL) in the surrounding pore water (see Fig. \ref{fig:EDL-layers}a). The EDL contains an excess of charge distributed in two layers that counterbalances the mineral one \citep[e.g.,][]{hunter1981zeta,revil2004streaming}: the Stern and the diffuse layer. When the water flows through the pore (Fig. \ref{fig:EDL-layers}c), the excess charge located in the diffuse layer (Fig. \ref{fig:EDL-layers}b) is dragged, which leads to a streaming current generation and a resulting electrical potential distribution. This EK phenomenon has been studied experimentally and theoretically for more than a century \citep[e.g.,][]{smoluchowski1903contribution}. While the generation of the streaming potential has been well studied and modeled under saturated conditions \citep[e.g.,][]{jardani2007tomography, revil2013self, guarracinophysically}, its generation under partially saturated conditions is still under discussion in the community and no consensus has been achieved regarding how to best model it.

Two main approaches can be used to model the streaming current generation under partially saturated conditions: (1) the Helmholtz-Smoluchowski coupling coefficient (or a variation of it to include the electrical surface conductivity) and (2) the excess charge that is effectively dragged in by the water. The streaming potential coupling coefficient has been defined to relate an electrical potential difference (i.e., the electrical field) and a hydraulic pressure difference (i.e., the groundwater flow). The first approach is therefore focused on the evolution of this parameter with varying water saturation \citep[e.g.,][]{guichet2003streaming, darnet2004modelling, revil2004streaming, jackson2010multiphase, vinogradov2011multiphase, allegre2015influence,fiorentino2016two}. The second approach results from a variable change where the coupling parameter is the excess charge that is effectively dragged by the water in the pore space \citep[e.g.,][]{linde2007streaming, revil2007electrokinetic, jougnot2012derivation, jougnot2015monitoring}. This approach allows the decomposition of the streaming potential coupling coefficient in three components: the relative permeability, the electrical conductivity, and the effective excess charge density. Each of them varies with saturation and can be determined independently. The fact that these components behave differently with saturation and strongly depend on the soil texture is a likely explanation of why the behavior of the streaming potential coupling coefficient cannot be expressed in general \citep[see discussion in][]{jougnot2012derivation}. While the variation of the relative permeability and the electrical conductivity under variably saturated conditions have been the subject of many works during the last decades, the variation of the effective excess charge is still largely understudied \citep[e.g.,][]{revil2007electrokinetic, jougnot2012derivation, jougnot2015monitoring, zhang2017streaming}. In this general context, the aim of our work is to develop a model following the excess charge approach that explicitly considers the dependence of the different parameters with saturation and the chemistry of the pore water.

Pore space in real porous media can be complex (e.g., the pores are three dimensional, some of them can be connected and have different geometrical shapes). Capillary tube models are a simple representation of the real pore space that have been used to provide valuable insight into transport in porous media. \cite{jackson2008characterization, jackson2010multiphase}, \cite{linde2009comment}, \cite{jougnot2012derivation,jougnot2015monitoring}, \cite{thanh2017fractal}, and \cite{guarracinophysically} have successfully used these models to study the streaming potential phenomenon. Fractal distributions of the capillary tubes has proven to be useful to characterize porous media when describing hydrological processes and hydraulic properties for different soil textures \citep[e.g.,][]{tyler1990fractal,yu2003permeabilities,ghanbarian2011review,guarracino2014fractal,xu2015discussion,soldi2017simple,thanh2017fractal}. In this study, we derive an analytical model to determine the effective excess charge density under partially saturated conditions. On the one hand, effective saturation and relative permeability curves are estimated by up-scaling the hydraulic properties at the pore scale to a bundle of capillary tubes with a fractal pore-size distribution. On the other hand, the effective excess charge density in a single capillary tube is calculated from the radial distributions of excess charge and water velocity. Therefore, at macroscopic scale the effective excess charge density is then estimated using a flux-averaging technique. Combining these hydraulic and EK properties, a closed-form expression for the effective excess charge density is obtained as a function of effective saturation and relative permeability. The resulting expression also relies on intrinsic petrophysical properties (i.e., permeability, porosity, tortuosity) and chemical interface properties (i.e., ionic concentration, zeta potential, Debye length). The performance of the proposed model is tested against different sets of experimental data and previous models. In addition, a relationship between the effective excess charge density and permeability can be derived. It is also shown that the estimates of the effective excess charge density can be satisfactorily extrapolated from saturated to partially saturated conditions by introducing the ratio of the effective saturation and relative permeability of the media.

\section{Streaming potential framework}
\label{sec:framework}

In the present section we introduce the theory used to describe the streaming current. Our work is based on the framework of \cite{revil2007electrokinetic} in which SP signals can be related to water velocity directly. It focuses on the excess charge approach proposed by \cite{jougnot2012derivation} that describes the generation of these currents as an excess charge effectively dragged by the water flow.

The SP response of a given source current density $\mathbf{J}_s$ (A/m$^2$) can be described by two equations \citep{sill1983self}:

\begin{equation}
\mathbf{J} = \sigma \mathbf{E} + \mathbf{J}_s,
\label{eq:sill-1J}
\end{equation}
\begin{equation}
\nabla \cdot \mathbf{J} = 0,
\label{eq:sill-2dot}
\end{equation}
where $\mathbf{J}$ (A/m$^2$) is the total current density, $\sigma$ (S/m) the bulk electrical conductivity, $\mathbf{E} = -\nabla \varphi $ (V/m) the electrical field being $\varphi$ (V) the electrical potential. In the absence of external current (i.e., no current injection in the medium), Eqs. \eqref{eq:sill-1J} and \eqref{eq:sill-2dot} can be combined to obtain:

\begin{equation}
\nabla \cdot \left( \sigma \nabla \varphi \right) = \nabla \cdot \mathbf{J}_s.
\label{eq:no-current}
\end{equation}
Since EK processes often dominates SP signals in hydrological applications, we consider the streaming current density $\mathbf{J}_s^{EK}$ as the source of $\mathbf{J}_s$. This current density is directly related to the water flux which follows Darcy's law \citep{darcy1856fontaines} and can be expressed as $\mathbf{u}$ (m/s):

\begin{equation}
\mathbf{u} = - \frac{k}{\eta_w} \mathbf{\nabla} (p - \rho_w g z) = -K \mathbf{\nabla} H
\label{eq:darcy-vel}
\end{equation}
where $k$ (m$^2$) is the permeability, $\eta_w$ (Pa s) the water dynamic viscosity, $p$ (Pa) the water pressure, $\rho_w$ (kg/m$^3$) the water density, $g$ (m/s$^2$) the gravitational acceleration, $K = \frac{\rho_w g k}{\eta_w}$ (m/s) the hydraulic conductivity , and $H = \frac{p}{\rho_w g} - z$ (m) the hydraulic head.

Traditionally this source current density is defined with the hydraulic gradient through the streaming potential coupling coefficient $C_{EK}$ (V/m) \citep{helmholtz1879studien, smoluchowski1903contribution}:

\begin{equation}
\mathbf{J}_s^{EK} = \sigma C_{EK} \mathbf{\nabla}H,
\label{eq:Jek}
\end{equation}
with $C_{EK}$ defined as:

\begin{equation}
C_{EK} = \dfrac{\partial \varphi}{\partial H} \Big |_{\mathbf{J} = \mathbf{0}},
\end{equation}
where this coefficient measures the sensitivity between the electrical potential and the variation of the hydraulic head of the water.

Based on simple analytical developments, \cite{kormiltsev1998three} and \cite{revil2004constitutive} express the coupling coefficient as a function of the effective excess charge density. Following Revil and collaborators formalism, it yields:

\begin{equation}
C_{EK}^{sat} = - \frac{\hat{Q}^{sat}_v k}{\sigma^{sat} \eta_w},
\label{eq:Cek-Q-sat}
\end{equation} 
where $\hat{Q}^{sat}_v$ (C/m$^3$) is the excess charge density in the diffuse layer per pore water volume and $\sigma^{sat}$ (S/m) the electrical conductivity of the medium at saturation. Equation \eqref{eq:Cek-Q-sat} can be extended under partially water-saturated conditions as follows \citep[e.g.,][]{linde2007streaming,jackson2010multiphase}:

\begin{equation}
C_{EK}(S_w) = - \frac{\hat{Q}_v(S_w)  \; k_{rel}(S_w)k}{\sigma(S_w) \; \eta_w},
\label{eq:Cek-Q-unsat}
\end{equation} 
where $k_{rel}$ is the relative permeability which is a dimensionless function of water saturation $S_w$ and varies between 0 and 1, $\sigma$ (S/m) the electrical conductivity of the media and $\hat{Q}_v$ (C/m$^3$) the excess charge density which are also functions of the water saturation. Several models from the literature can be used to describe the dependence of these petrophysical properties $k_{rel}$ and $\sigma$ with $S_w$. Two of the most widely used model to estimate $k_{rel}$ are \cite{brooks1964hydraulic} and \cite{van1980closed}, while Archie's second law (\citeyear{archie1942electrical}) and \cite{waxman1968electrical} are commonly used to predict the electrical conductivity.

Note that $\mathbf{J}_s^{EK}$ (Eq. \ref{eq:Jek}) can be expressed using directly the effective excess charge density under partially saturated conditions and the flux velocity, it yields:

\begin{equation}
\mathbf{J}_s^{EK} = \hat{Q}_v(S_w) \mathbf{u}
\label{eq:Jek(Q)}
\end{equation} 

In the following subsections the hydraulic and electrokinetic properties at microscale (for one single pore) and macroscale (representative elementary volume, REV) scale are presented.

\subsection{Pore scale}
\label{sec:porescale}

\subsubsection{Hydraulic properties}
\label{sec:porehydprop}

The porous medium is represented by a bundle of circular-tortuous capillary tubes. Each pore is conceptualized as a cylindrical tube of radius $R$ (m), length $l$ (m) and hydraulic tortuosity $\tau$ (dimensionless) which can be defined as the ratio $l/L$ being $L$ (m) the length of the REV at macroscale.

The volume of a single pore is then given by:

\begin{equation}
V_p (R)= \pi R^2 l,
\label{eq:vol-pore}
\end{equation}
and the average velocity in the capillary tube $\bar{v}$ (m/s) can be obtained from the Poiseuille law, assuming laminar flow, as:

\begin{equation}
\bar{v}(R)= \frac{\rho_w g}{8 \eta_w \tau} R^2 \frac{\Delta h}{L},
\label{eq:average-vel}
\end{equation}
where $\Delta h$ (m) is the pressure head drop across the REV.

\subsubsection{Electrokinetic properties}
\label{sec:poreekprop}

\begin{figure}[ht]
\centering
\includegraphics[width=.5 \textwidth,keepaspectratio=true]{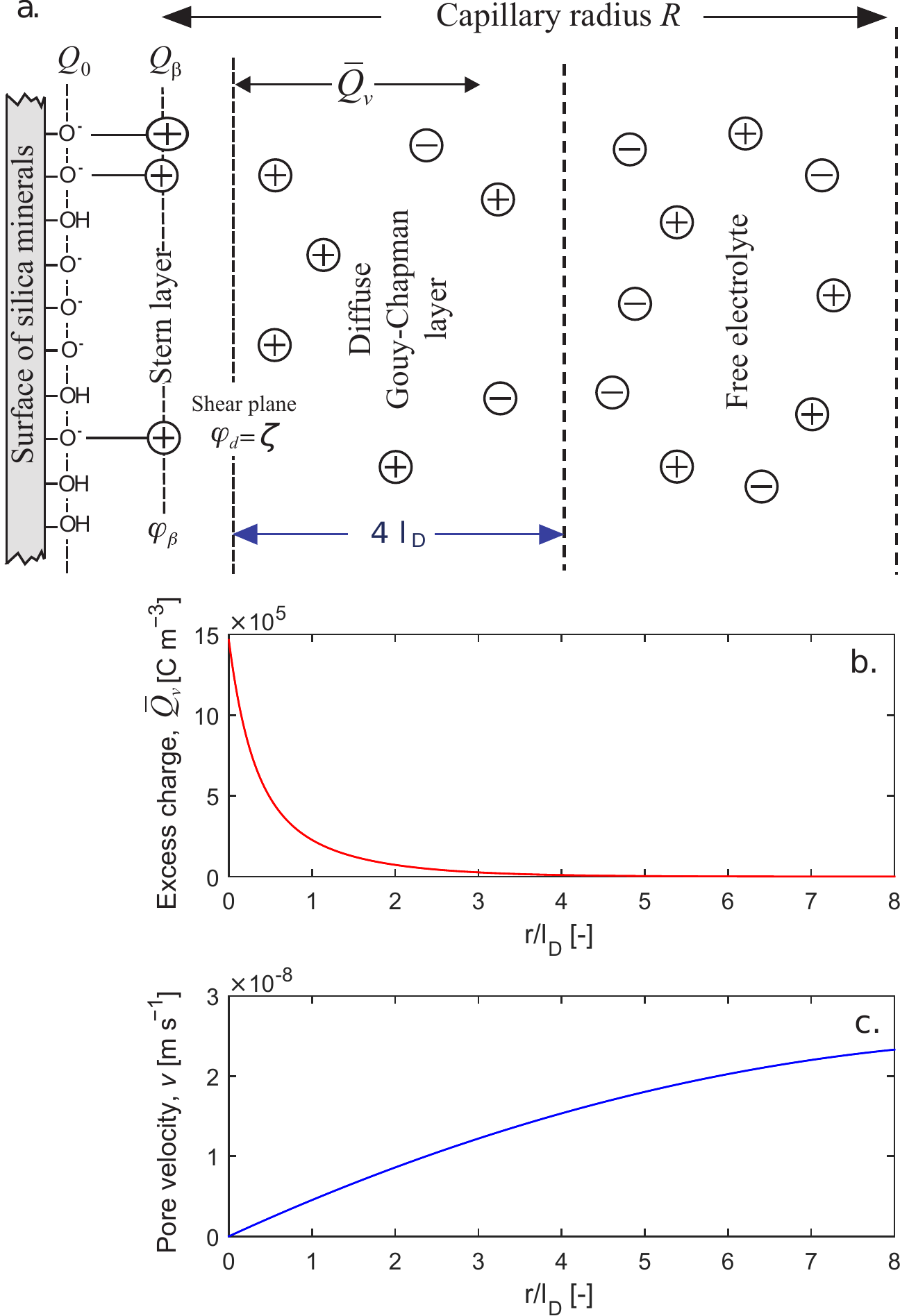}
\caption{(a) Scheme of the electrical layers within a capillary tube of radius R. (b) and (c) distributions of the static excess charge and the pore water velocity as a function of the distance from the mineral surface in the capillary, respectively.} \label{fig:EDL-layers}
\end{figure}

The electrokinetic behavior of a capillary is mainly determined by the previously derived $\bar{v}(R)$ and the chemical composition of the pore water. Then, we consider that each capillary tube is saturated by a binary symmetric 1:1 electrolyte with a ionic concentration $C_w^0$ (mol/L) (e.g., NaCl). Since the mineral's surface is generally electrically charged, there exists an excess of charge in the pore water in order to assure the electrical neutrality: the electrical double layer (EDL). It can be distinguished from the free electrolyte where there are no charges present (see Fig. \ref{fig:EDL-layers}). The EDL is composed by two layers, close to the capillary wall is where most of that excess charge is fixed in the Stern layer while the remaining part of it is distributed in the Gouy-Chapman layer (or diffuse layer, Fig. \ref{fig:EDL-layers}a,b). The interface between these two layers can be approximated by a plane of shear which separates the stationary and non-stationary fluid \citep{hunter1981zeta}. This plane is characterized by an electrical potential called zeta potential $\zeta$ (V), which depends on ionic strength, temperature, pH, among other quantities \citep[e.g.,][]{revil1999streaming}.

To study the electrokinetic properties, the Stern layer can be neglected since it is beyond the shear plane and the water flow \citep[e.g.,][]{leroy2004triple, tournassat2009comparison}. Therefore, we are interested in the Gouy-Chapman layer where the fluid is non-stationary, the thickness of this layer is given in terms of the Debye length $l_D$ (m) which is defined as \citep{hunter1981zeta}:

\begin{equation}
l_D = \sqrt{\frac{\epsilon k_B T}{2 N_A C_w^0 e_0^2}}
\label{eq:debyelength}
\end{equation}
where $\epsilon$ (F/m) is the pore water dielectric permittivity, $k_B$ (J/K) the Boltzmann constant, $T$ (K) the absolute temperature, $N_A$ (1/mol) the Avogadro's number, $C^0_w$ the ionic concentration far from the mineral surface and $e_0$ (C) the elementary charge. The Gouy-Chapman layer's (i.e., diffuse layer) thickness is assumed to be four Debye lengths \citep{hunter1981zeta}. Indeed, as it can be seen in Fig. \ref{fig:EDL-layers}, the excess charge distribution tends to zero as the distance to the mineral surface diminishes.

Recently, \cite{guarracinophysically} developed a physically based analytical model to estimate the effective excess charge density dragged in the capillary by the water flow under saturated conditions (see Fig. \ref{fig:EDL-layers}b and c). Considering the thin double layer assumption (i.e., the thickness of the double layer is small compared to the pore size, $l_D \ll R$), they found that the effective excess charge density carried by the water flow in a single tube of radius $R$ is given by:

\begin{equation}
\hat{Q}_v^R = \frac{8 N_A e_0 C^0_w}{(R/l_D)^2} \left[- \frac{2 e_0 \zeta}{k_B T} - \left( \frac{e_0 \zeta}{3 k_B T} \right)^3 \right].
\label{eq:Qv-sat-pore}
\end{equation}
Note that according to \cite{guarracinophysically}, the above equation is valid for capillaries whose radii $R$ are greater than $5 l_D$. In Section \ref{subsec:ek-prop}, Eq. \eqref{eq:Qv-sat-pore} is used to estimate $\hat{Q}_v$ in a porous medium under partially saturated conditions.

\subsection{REV scale}
\label{sec:rev}

In order to derive the hydraulic and electrokinetic properties at macroscale we consider as representative elementary volume (REV) a cylinder of radius $R_{REV}$ (m) and length $L$ (m). The porous medium of the REV is conceptualized as an equivalent bundle of capillary tubes with a fractal pore size distribution and the pore structure is represented by the geometry described in the previous section with radii varying from a minimum pore radius $R_{min}$ (m) to a maximum pore radius $R_{max}$ (m).

The cumulative size-distribution of pores whose average radii are greater than or equal to $R$ is assumed to obey the following fractal law \citep{tyler1990fractal, yu2003permeabilities, guarracino2014fractal, soldi2017simple,guarracinophysically}:

\begin{equation}
N(R)= \left( \frac{R_{REV}}{R} \right)^D
\label{eq:fractal-law}
\end{equation}
where $D$ is the fractal dimension of pore size with $1 < D < 2$ and $0 < R_{min} \leqslant R \leqslant R_{max} < R_{REV}$. It is worth mentioning that $D$ can be considered as a measure of the soil texture where the highest values of $D$ are associated with the finest textured soils \citep{tyler1990fractal}. Differentiating Eq. \eqref{eq:fractal-law} with respect to $R$ we obtain the number of pores whose radii are in the infinitesimal range $R$ to $R + dR$:

\begin{equation}
dN(R)= -D R_{REV}^D R^{-D-1}dR
\label{eq:dN}
\end{equation}
where the negative sign implies that the number of pores decreases with the increase of pore radius $R$.

The porosity of the REV can be computed from its definition as the quotient between the volume of pores and the volume of the REV which yields \citep{guarracinophysically}:

\begin{equation}
\phi= \frac{\tau D}{R_{REV}^{2-D}(2-D)} \left( R_{max}^{2-D} - R_{min}^{2-D}\right).
\label{eq:porosity}
\end{equation}
The parameter $\tau$ can be defined for the entire model because the length of the pores is assumed to be independent of the capillary radius (i.e., $\tau$ is the same for all the capillaries).

\subsubsection{Saturation and relative permeability curves}
\label{sec:curves}
In this section, we consider the REV under partially saturated conditions. Then, the contribution to the water flow is given by the effective saturation $S_e$ (dimensionless), defined by: 

\begin{equation}
S_e = \frac{S_w - S_w^r}{1 - S_w^r},
\label{eq:Se-definition}
\end{equation}
where $S_w$  and $S_w^r$ are the water saturation and residual water saturation, respectively.

In order to obtain the effective saturation curve, we consider that the REV is initially fully saturated and then drained when submitted to a pressure head $h$ (m). For a straight capillary tube, we can relate a pore radius $R_h$ (m) to $h$ by the following equation \citep[e.g.,][]{jurin1717account, bear1998dynamics}:

\begin{equation}
h = \frac{2 T_s \cos (\beta)}{\rho_w g R_h},
\label{eq:h}
\end{equation}
where $T_s$ (N/m) is the surface tension of the water and $\beta$, the contact angle. A capillary becomes fully desaturated if its radius $R$ is greater than the radius $R_h$ given by Eq. \eqref{eq:h}. Then, it is reasonable to assume that pores with radii $R$ between $R_{min}$ and $R_h$ will remain fully saturated. Therefore, according to Eqs. \eqref{eq:vol-pore} and \eqref{eq:dN}, the effective saturation curve $S_e$ can be computed by:

\begin{equation}
S_e = \frac{\int_{R_{min}}^{R_h} V_p(R) dN}{\int_{R_{min}}^{R_{max}} V_p(R) dN} = \frac{R_h^{2-D} - R_{min}^{2-D}}{R_{max}^{2-D} -R_{min}^{2-D} }.
\label{eq:sat}
\end{equation}
Using the same hypothesis and neglecting film flow on capillary surfaces, we can obtain the relative permeability curve. The main contribution to the volumetric flow through the REV $q$ (m$^3$/s) can be computed by integrating the individual volumetric flow rates over the pores that remain fully saturated ($R_{min} \leq R \leq R_h$):

\begin{equation}
q = \int_{R_{min}}^{R_h} \bar{v}(R) \pi R^2 dN = \frac{\rho_w g}{8 \eta_w \tau} \frac{\Delta h}{L} \frac{\pi D R_{REV}^D}{4-D} \left(R_h^{4-D}-R_{min}^{4-D}\right).
\label{eq:flow-unsat}
\end{equation}
Otherwise, according to Buckingham-Darcy's law (\citeyear{buckingham1907studies}), the total volumetric flow rate through the REV can be expressed as:

\begin{equation}
q = \frac{\rho_w g}{\eta_w} k k_{rel} \frac{\Delta h}{L} \pi R_{REV}^2.
\label{eq:buck-darcy}
\end{equation}
Combining Eqs. \eqref{eq:flow-unsat} and \eqref{eq:buck-darcy} we can define analytical expressions for permeability and relative permeability:

\begin{equation}
k = \frac{D}{8 \tau (4-D) R_{REV}^{2-D}} \left(R_{max}^{4-D}-R_{min}^{4-D}\right),
\label{eq:k-sat}
\end{equation}
and

\begin{equation}
k_{rel}(R_h) = \frac{R_h^{4-D}-R_{min}^{4-D}}{R_{max}^{4-D}-R_{min}^{4-D}}.
\label{eq:kr}
\end{equation}

Relative permeability $k_{rel}$ can also be expressed in terms of effective saturation $S_e$. By combining Eqs. \eqref{eq:sat} and \eqref{eq:kr} we obtain the following equation:

\begin{equation}
k_{rel} = \frac{ \left[ S_e \left( 1-  \alpha^{2-D} \right) + \alpha^{2-D} \right]^\frac{4-D}{2-D} - \alpha^{4-D} }{ 1- \alpha^{4-D} },
\label{eq:kdeSyRmin-max}
\end{equation}
where

\begin{equation}
\alpha = \frac{R_{min}}{R_{max}}.
\label{eq:alpha}
\end{equation}
Note that the parameter $\alpha$ can be used as a measurement of the soil gradation. High values of $\alpha$ can be associated to well graded soils while low values to poorly graded soils. Then, for $R_{max} \gg R_{min}$ ($\alpha \rightarrow 0$), Eq. \eqref{eq:kdeSyRmin-max} can be reduced to: 

\begin{equation}
k_{rel}(S_e) = S_e^{\frac{4-D}{2-D}}.
\label{eq:k(S)}
\end{equation}
It can be observed that this expression is similar to the power law of the well-known Brooks and Corey model (\citeyear{brooks1964hydraulic}).

\subsubsection{Electrokinetic properties}
\label{subsec:ek-prop}

The electrokinetic phenomenon is a coupling between hydraulic and electrical processes in a medium. Based on the previous description of macroscopic hydraulic properties we compute the effective excess charge density $\hat{Q}_v^{REV}$ carried by the water flow in the REV. We consider similar conditions of saturation as in Section \ref{sec:curves}: the REV is initially fully saturated and a pressure head $h$ is applied, drying the larger pores. Therefore, only the capillaries that remain fully saturated ($R_{min} \leq R \leq R_h$) contribute to the effective excess charge density $\hat{Q}_v^{REV}$ (C/m$^3$) that can be computed by:

\begin{equation}
\hat{Q}_v^{REV} = \frac{1}{v_D \pi R_{REV}^2} \int_{R_{min}}^{R_h} \hat{Q}_v^R \bar{v}(R) \pi R^2 dN,
\label{eq:Qv-unsat-def}
\end{equation}
where $v_D = \frac{\rho_w g}{\eta_w} k_{rel} k \frac{\Delta h}{L}$ (m/s) is the Darcy's flow. Substituting Eqs. \eqref{eq:average-vel}, \eqref{eq:Qv-sat-pore} and \eqref{eq:dN} in Eq. \eqref{eq:Qv-unsat-def} yields:

\begin{equation}
\hat{Q}_v^{REV} = 8 N_A e_0 C^0_w l_D^2 \left[- \frac{2 e_0 \zeta}{k_B T} - \left( \frac{e_0 \zeta}{3 k_B T} \right)^3 \right] \frac{\rho_w g}{8 \eta_w \tau} \frac{\Delta h}{L} \frac{D}{v_D R_{REV}^{2-D}(2-D)} \left[R_h^{2-D} - R_{min}^{2-D} \right].
\label{eq:Qv-unsat}
\end{equation}
Finally, combining Eqs. \eqref{eq:porosity}, \eqref{eq:sat} and \eqref{eq:Qv-unsat} we obtain the following expression for $\hat{Q}_v^{REV}$:

\begin{equation}
\hat{Q}_v^{REV}(S_e,C^0_w) = N_A e_0 C^0_w\left[- \frac{2 e_0 \zeta}{k_B T} - \left( \frac{e_0 \zeta}{3 k_B T} \right)^3 \right]\left(\frac{l_D}{\tau}\right)^2 \frac{\phi}{k} \frac{S_e}{k_{rel}(S_e)}.
\label{eq:Qv-proposedmodel}
\end{equation}
This equation is the main contribution of this paper as it describes the entire model to determine the effective excess charge density in a porous medium under partially saturated conditions derived from geometrical properties and physical laws. This closed-form expression relies on two kind of medium properties: (1) petrophysical properties, i.e., permeability, porosity, effective saturation and hydraulic tortuosity, and (2) electro-chemical properties, i.e., ionic concentration, zeta potential, and Debye length.

\begin{figure}
\includegraphics[width=1\textwidth,keepaspectratio=true]{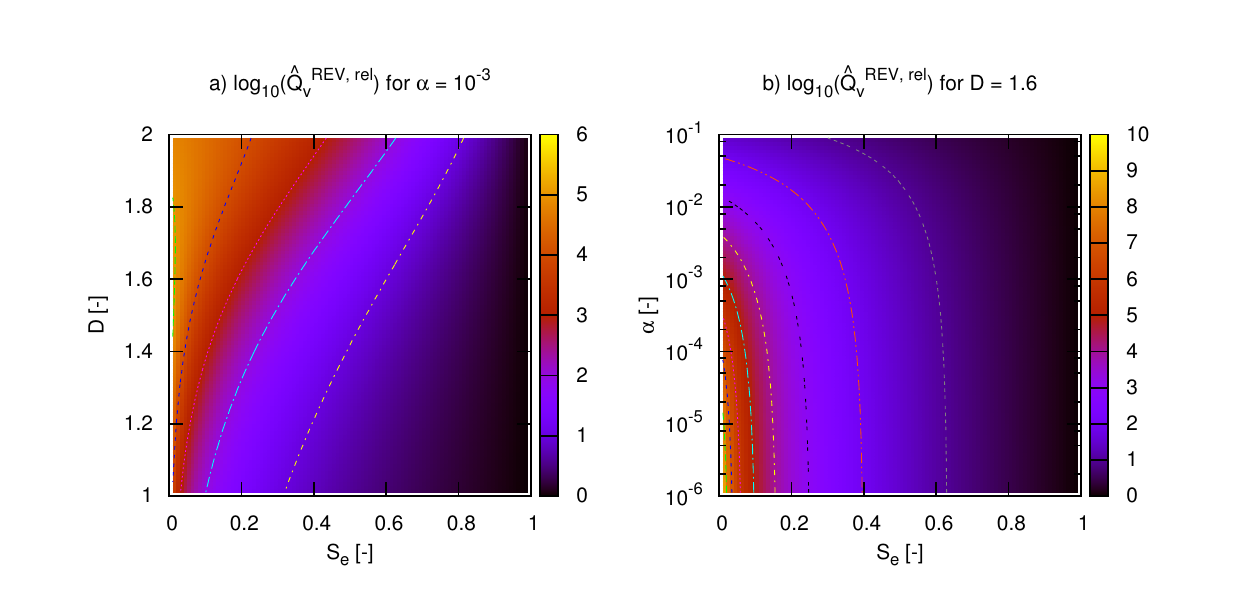}
\caption{Parametric analysis of relative effective excess charge density $\hat{Q}_v^{REV,rel}$ as a function of effective saturation $S_e$: a) sensitivity to fractal dimension $D$ for a reference value of $\alpha$, and b) sensitivity to $\alpha$ for a reference value of $D$}
\label{fig:parameters}
\end{figure}

Note that under saturated conditions (i.e., $S_e=1$), Eq. \eqref{eq:Qv-proposedmodel} is equivalent to the model of \cite{guarracinophysically}. Then, $\hat{Q}_v^{REV}$ can be expressed as the product between the effective excess charge density for saturated conditions $\hat{Q}_v^{REV,sat}$ (C/m$^3$) and the relative effective excess charge density $\hat{Q}_v^{REV,rel}$ \citep[dimensionless, see][]{jougnot2015monitoring}:

\begin{equation}
\hat{Q}_v^{REV} = \hat{Q}_v^{REV,sat} \hat{Q}_v^{REV,rel}(S_e)
\label{eq:Qsat}
\end{equation}
where

\begin{equation}
\hat{Q}_v^{REV,rel}(S_e) = \frac{S_e}{k_{rel}(S_e)}.
\label{eq:Qrel-generic}
\end{equation}
It can be seen that the relative effective excess charge density defined by Eq. \eqref{eq:Qrel-generic} depends only on hydraulic variables. By considering the proposed relationship between $k_{rel}$ and $S_e$ (Eq. \eqref{eq:kdeSyRmin-max}), the following analytical expression of Eq. \eqref{eq:Qrel-generic} can be obtained:

\begin{equation}
\hat{Q}_v^{REV,rel}(S_e) = \frac{S_e (\alpha^{D-4}-1)}{\left[S_e \left(\alpha^{D-2}-1\right)+1\right]^\frac{4-D}{2-D}-1}.
\label{eq:Qrel-kprop}
\end{equation}
The above equation depends on the model parameters $D$ and $\alpha$, to test their role on $\hat{Q}_v^{REV,rel}$ estimates, we perform a parametric analysis. Figure \ref{fig:parameters} shows log($\hat{Q}_v^{REV,rel}$) values for the following ranges of variability of each parameter: $1 < D < 2$ and $10^{-6} < \alpha < 10^{-1}$ while we considered as reference values $D = 1.6$ and $\alpha = 10^{-3}$. Figure \ref{fig:parameters}a shows that the relative effective excess charge density $\hat{Q}_v^{REV,rel}$ decreases faster when effective saturation decreases for low fractal dimension values which are associated to coarse textured soils. Figure \ref{fig:parameters}b shows that for low values of parameter $\alpha$ (e.g. well graded soils), $\hat{Q}_v^{REV,rel}$ decreases faster and its values change in 10 orders of magnitude. By comparing both panels, it can be noticed that $\hat{Q}_v^{REV,rel}$ is greater than 1 and may vary several orders of magnitude. Indeed, variations of parameter $\alpha$ produce the more significant effect in $\hat{Q}_v^{REV,rel}$ values.

A similar expression for the relative effective excess charge density was obtained by \cite{jackson2010multiphase} based on the coupling coefficient approach to estimate the streaming currents in a bundle of capillary tubes (see more details in Sec. \ref{subsec:previous-models}). Nevertheless, it is important to remark that the proposed model provides an analytical closed-form expression (Eq. \eqref{eq:Qv-proposedmodel}) to estimate the total effective excess charge density $\hat{Q}_v^{REV}$.

Assuming that $R_{min} \ll R_{max}$, Eq. \eqref{eq:Qrel-kprop} can be expressed as:

\begin{equation}
\hat{Q}_v^{REV,rel}(S_e) = \frac{1}{S_e^{a}}.
\label{eq:Qrel-proposed}
\end{equation}
where $a = 2/(2-D)$. Note that $\hat{Q}_v^{REV,rel}$ depends inversely on effective saturation with exponent $a > 2$. \cite{linde2007streaming} proposed a volume averaging model to estimate $\hat{Q}_v^{REV,rel}$ as a function of water saturation which considers $a=1$ (more details in Sec. \ref{subsec:previous-models}). However, this model has been shown to underestimate the effective excess charge when compared to flux averaging approaches \citep[e.g.,][]{jougnot2012derivation}.

\subsection{Relationship between $\hat{Q}_v^{REV}$ and permeability}
\label{sec:jardani}

In this section, we derive an expression to estimate the effective excess charge density $\hat{Q}_v^{REV}$ from permeability under unsaturated conditions. Assuming that $R_{min} \ll R_{max}$, a relationship between the petrophysical properties of the medium $\phi$ and $k$ can be obtained by combining Eqs. \eqref{eq:porosity} and \eqref{eq:k-sat}:

\begin{equation}
\phi = {\left(\frac{k}{\gamma}\right)}^{\frac{2-D}{4-D}},
\label{eq:kozeny-proposed}
\end{equation}
where $\gamma = \frac{D R^2_{REV}}{8 \tau (4-D)}\left(\frac{2-D}{\tau D}\right)^{\frac{4-D}{2-D}}$. Replacing this expression in Eq.\eqref{eq:Qv-proposedmodel} and taking logarithm on both sides of the resulting equation, it yields

\begin{equation}
\text{log}_{10}\left(\hat{Q}_v^{REV}\right) = A + B\, \text{log}_{10}(k \; k_{rel}),
\label{eq:jardani-unsat}
\end{equation}
where

\begin{equation}
A = \text{log}_{10} \left\lbrace \frac{N_A e_0 C_w^0}{\gamma^\frac{2-D}{4-D}} \left[ -\frac{2 e_0 \zeta}{k_B T} -\left(\frac{ e_0 \zeta}{3 k_B T}\right)^3\right] \left(\frac{l_D}{\tau}\right)^2 \right\rbrace ,
\end{equation}
and

\begin{equation}
B = -\frac{2}{4-D}.
\end{equation}
Note that the constant $A$ depends on chemical and hydraulic parameters, while constant $B$ depends only on the fractal dimension. One can remark that under saturated conditions, $k_{rel}=1$, Eq. \eqref{eq:jardani-unsat} is equivalent to the empirical relationship proposed by \cite{jardani2007tomography} where $A = -9.2349$ and $B = -0.8219$. These constant values were obtained by fitting the log($\hat{Q}_v^{REV,sat}$)-log($k$) relationship to a large set of experimental data including different lithologies and salinities. This empirical relationship has been successfully used in several studies to directly estimate $\hat{Q}_v^{REV,sat}$ values from permeability \citep[e.g.,][]{revil2013coupled, jardani2009stochastic,jougnot2013seismoelectric, roubinet2016streaming}. Besides, Eq. \eqref{eq:jardani-unsat} represents the extension to partially saturated media of the recent analytical expression derived by \cite{guarracinophysically} which provides a theoretical justification to the relationship of \cite{jardani2007tomography}. 

\begin{figure}
\centering
\includegraphics[width=0.8\textwidth,keepaspectratio=true]{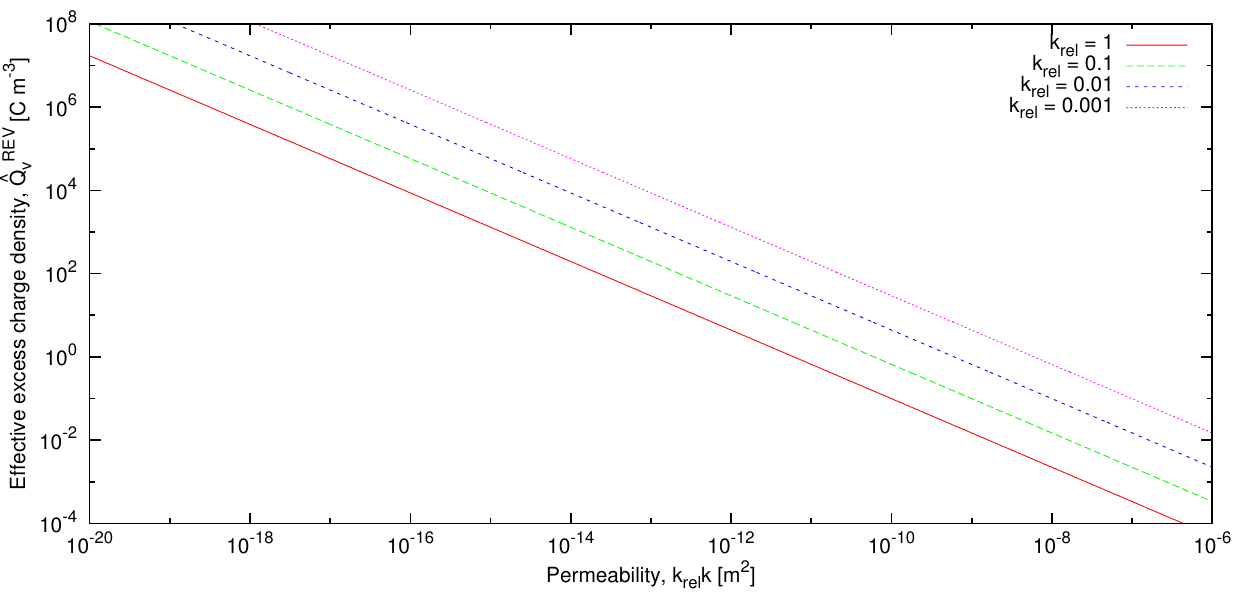}
\caption{Effective excess charge density as a function of the product between the relative and intrinsic permeabilities. The red solid line represents the empirical relationship for saturated conditions of \cite{jardani2007tomography}.}
\label{fig:jardani}
\end{figure}

Figure \ref{fig:jardani} shows the effective excess charge density $\hat{Q}_v^{REV}$ as a function of permeability for different $k_{rel}$ values. The empirical relationship proposed by \cite{jardani2007tomography} is also shown in the Figure. Note that while the slope of the curves remains unchanged, the resulting estimates of $\hat{Q}_v^{REV}$ are displaced to higher values when $k_{rel}$ decreases.

\section{Application of the model to different hydrodynamic properties}
\label{sec:sensitivity}
In this section, we analyze the effect of soil texture and permeability models on the estimates of the relative effective excess charge density.

\subsection{Estimation of $\hat{Q}_v^{REV,rel}$ for different soil textures}
\label{subsec:textures}

The relative effective excess charge density depends only on the hydraulic properties of the medium, $S_e$ and $k_{rel}$. Hence, to study the effect of soil textures on $\hat{Q}_v^{REV,rel}$, we consider that the proposed permeability model is related to soil textures through the fractal dimension $D$. Given the similarity between the proposed and Brooks and Corey (\citeyear{brooks1964hydraulic}) models, we compare both $S_e(h)$ expressions to obtain a relationship between their parameters. Therefore, the performance of Eq. \eqref{eq:Qrel-proposed} to estimate $\hat{Q}_v^{REV,rel}$ can be analyzed using typical values of Brooks and Corey parameters for 11 soil textures from \cite{brakensiek1992comment} (see Table \ref{table:BC}). Assuming $R_{max} \gg R_{min}$, Eq. \eqref{eq:sat} can be expressed as:

\begin{equation}
S_e = \left( \frac{R_h}{R_{max}} \right) ^{2-D} = \left( \frac{h_{min}}{h} \right) ^{2-D}
\label{eq:saturation-proposed}
\end{equation}
where $h_{min}$ is obtained using Eq. \eqref{eq:h}.

The water retention equation proposed by Brooks and Corey model (\citeyear{brooks1964hydraulic}) is given by:

\begin{equation}
S_e = \left( \frac{h_b}{h} \right) ^\lambda
\label{eq:saturation-BC}
\end{equation}
where $\lambda$ is the pore-size distribution index and $h_b$ the so called bubbling pressure head. Comparing Eqs. \eqref{eq:saturation-proposed} and \eqref{eq:saturation-BC}, the model parameters can be related through $D = 2 - \lambda$ and $R_{max} = \frac{2 T_s}{\rho_w g h_b}$ where the contact angle $\beta$ is assumed to be zero \citep{bear1998dynamics}. Table \ref{table:BC} lists the resulting parameters for all the soil textures.

\begin{figure}
\centering
\includegraphics[width=0.8\textwidth,keepaspectratio=true]{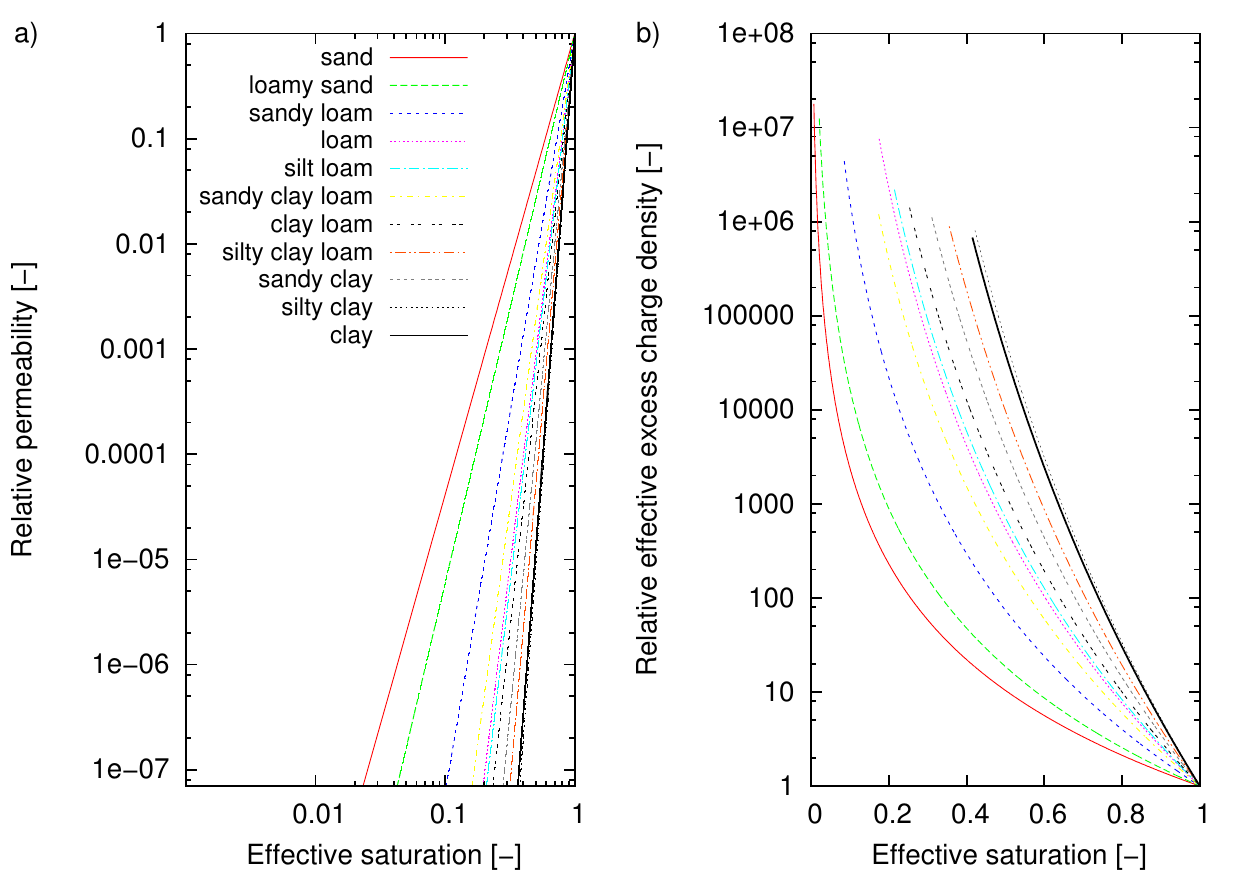}
\caption{Analysis of the proposed model for different soil textures. a) The relative permeability is calculated using Eq. \eqref{eq:k(S)} and b) the relative effective excess charge density is estimated using Eq. \eqref{eq:Qrel-proposed}}
\label{fig:textures}
\end{figure}

In Figure \ref{fig:textures} values of $k_{rel}$ and $\hat{Q}_v^{REV,rel}$ are presented as a function of $S_e$ assuming a ionic concentration of 1$\times10^{-3}$ mol/L. The effective saturation values are limited to the values predicted by Eq. \eqref{eq:saturation-proposed} within the range $5 l_D \leq R \leq R_{max}$, that is the radius range where the electrical double layer from the pore walls do not overlap and Eq. \eqref{eq:Qv-sat-pore} is valid \citep{guarracinophysically}. It is interesting to observe that for a fixed saturation value, the finest soil textures produce the higher values of $\hat{Q}_v^{REV,rel}$ (e.g. for $S_e = 0.6$, the $\hat{Q}_v^{REV,rel}$ values for sand and clay differ in 3 orders of magnitude). Also note that the maximum values of relative effective excess charge density are in the range of $10^6 - 10^7$ for all the soil textures being the higher values associated to fine soil textures at high $S_e$ values.

\begin{table}
\centering
\caption{Brooks and Corey parameters from the water retention equation for different soil textures \citep{brakensiek1992comment} and the corresponding values of the proposed model parameters}
\label{table:BC}
\begin{tabular}[t]{lcccc}

\hline\noalign{\smallskip}
Texture  & \multicolumn{2}{c}{Brooks and Corey} & \multicolumn{2}{c}{Proposed model} \\
 & $h_b [m] $ & $\lambda $ & $R_{max} [m] $ & $D $\\
\noalign{\smallskip}\hline\noalign{\smallskip}
Sand & 0.073 & 0.592 & 2.044$\times 10^{-4}$ & 1.408 \\
Loamy sand & 0.087 & 0.474 & 1.707$\times 10^{-4}$ & 1.526 \\
Sandy loam & 0.147 & 0.322 & 1.012$\times 10^{-4}$ & 1.678 \\
Loam & 0.112 & 0.220  & 1.329$\times 10^{-4}$ & 1.780 \\
Silt loam & 0.208 & 0.211 & 7.147$\times 10^{-5}$ & 1.789 \\
Sandy clay loam & 0.281 & 0.250 & 5.284$\times 10^{-5}$ & 1.750 \\
Clay loam & 0.259 & 0.194 & 5.731$\times 10^{-5}$ & 1.806 \\
Silty clay loam & 0.326 & 0.151 & 4.557$\times 10^{-5}$ & 1.849 \\
Sandy clay & 0.292 & 0.168 & 5.086$\times 10^{-5}$ & 1.832 \\
Silty clay & 0.342 & 0.127 & 4.339$\times 10^{-5}$ & 1.873 \\
Clay & 0.373 & 0.131 & 3.978$\times 10^{-5}$ & 1.869 \\
\noalign{\smallskip}\hline
\end{tabular}
\end{table}

\subsection{Estimation of $\hat{Q}_v^{REV,rel}$ using different constitutive models}
\label{subsec:Qrel-krel}
In order to test the performance of $k_{rel}$ models to estimate $\hat{Q}_v^{REV,rel}$, we fit predicted relative effective excess charge density values calculated as $S_e/k_{rel}$ with \cite{van1980closed} and \cite{brooks1964hydraulic} models, and the proposed $k_{rel}(S_e)$ relationship (Eq. \eqref{eq:Qrel-kprop}). For this analysis, we considered measured effective saturation-relative permeability data from \cite{mualem1976catalogue}. Then, an analytical expression for $\hat{Q}_v^{REV,rel}$ can be obtained for each relative permeability model:

\begin{equation}
\hat{Q}_v^{REV,rel} = \frac{S_e^{1/2}}{\left[1-\left(1-S_e^{1/m}\right)^{m}\right]^2},
\label{eq:Qrel-vG}
\end{equation}
\begin{equation}
\hat{Q}_v^{REV,rel} = \frac{1}{S_e^{2+2/\lambda}},
\label{eq:Qrel-BC}
\end{equation}
where $m$ and $\lambda$ are dimensionless empirical parameters related to the pore size distribution in the medium for the van Genuchten and Brooks and Corey models, respectively.

\begin{figure}
\centering
\includegraphics[width=0.8\textwidth,keepaspectratio=true]{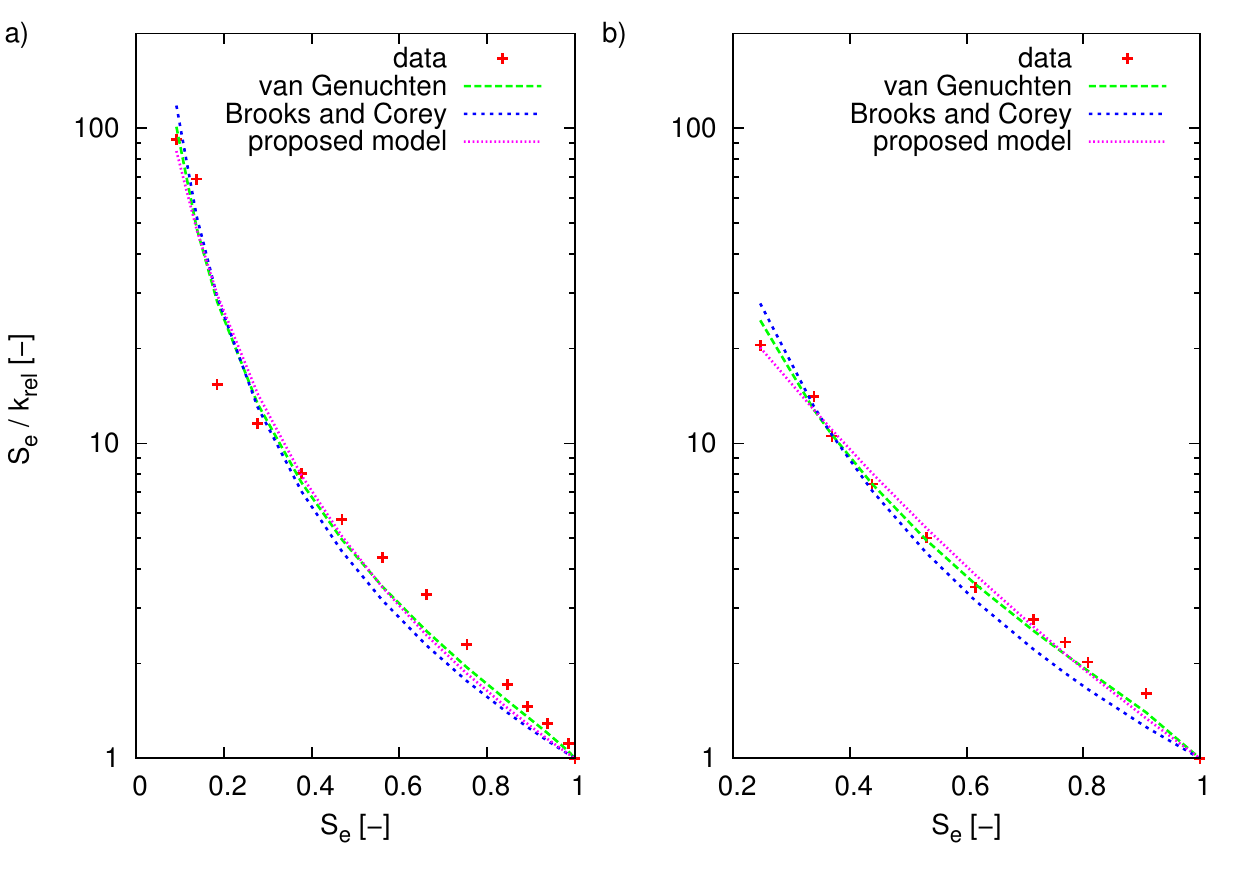}
\caption{Analysis of $\hat{Q}_v^{REV,rel}$ for different $k_{rel}$ models (van Genuchten, Brooks and Corey and Eq. \eqref{eq:kdeSyRmin-max}) for two soils: a) Sable de riviere and b) Rubicon sandy loam (data from \cite{mualem1976catalogue})}
\label{fig:testkrel}
\end{figure}

The parameters of the different $k_{rel}$ models ($m$, $\lambda$, $D$ and $\alpha$) were fitted using an exhaustive search method and the root-mean-square deviation (RMSD) was calculated taking logarithm differences due to the wide range of $\hat{Q}_v^{REV,rel}$ values. Table \ref{table:Qrel-krelmodels} lists the resulting parameter values and Figure \ref{fig:testkrel} shows the comparison between the fit of Eqs. \eqref{eq:Qrel-kprop}, \eqref{eq:Qrel-vG} and \eqref{eq:Qrel-BC} to two sets of experimental data (Rubicon sandy loam and Sable de riviere). It can be noticed that all the models produce good agreements with the predicted data. However, for Sable de riviere, parameter $\lambda$ requires unrealistic values in order to fit $\hat{Q}_v^{REV,rel}$ data at low saturations.

\begin{table}
\centering
\caption{Values of the fitted parameters for van Genuchten (\citeyear{van1980closed}), Brooks and Corey (\citeyear{brooks1964hydraulic}) and the proposed relative permeability model (Eq. \eqref{eq:kdeSyRmin-max}), and the corresponding RMSD}
\label{table:Qrel-krelmodels}
\begin{tabular}[t]{lccccccc}
\hline\noalign{\smallskip}
Soil type  & \multicolumn{2}{c}{van Genuchten} & \multicolumn{2}{c}{Brooks and Corey} & \multicolumn{3}{c}{Eq. \eqref{eq:kdeSyRmin-max}} \\
 & $m$ & RMSD & $\lambda^{-1}$& RMSD & $D$ & $\alpha$ & RMSD \\
\noalign{\smallskip}\hline\noalign{\smallskip}
Sable de riviere & 0.863 & 0.226 & 0 & 0.270 & 1.154 & 0.025 & 0.250 \\
Rubicon sandy loam & 0.802 & 0.086 & 0.186 & 0.171 & 1.597 & 0.057 & 0.086\\
\noalign{\smallskip}\hline
\end{tabular}
\end{table}

\section{Previous models}
\label{subsec:previous-models}
In this section, we review some recent models to describe SP signals generated under partially saturated conditions (see Table \ref{table:models}), based on both the coupling coefficient and the effective excess charge approaches.


\begin{table}
\centering
\caption{List of some of the previous models describing the electrokinetic phenomenon in SP signals derived from the coupling coefficient or excess charge approach}
\label{table:models}
\begin{tabular}[t]{lcc}
\hline\noalign{\smallskip}
Reference & Coupling coefficient approach & Excess charge approach \\
\noalign{\smallskip}\hline\noalign{\smallskip}

\begin{minipage}{3cm}
\cite{wurmstich1994modeling}$^{(1)}$
\end{minipage}  &
\parbox{4cm}{\begin{equation*}
C_{EK} = \frac{(1-w)C_v}{S_w^n}
\end{equation*}} & 
\parbox{7cm}{\begin{equation*}
-
\end{equation*}}
\\

\begin{minipage}{2.8cm}
\cite{perrier2000characterization}
\end{minipage}  & 
\parbox{4cm}{\begin{equation*}
C_{EK} = \frac{C_{EK}^{sat} k_{rel}(S_w)}{S_w^n}
\end{equation*}} &
\parbox{7cm}{\begin{equation*}
-
\end{equation*}} \\

\begin{minipage}{3cm}
\cite{guichet2003streaming}
\end{minipage}  & 
\parbox{4cm}{\begin{equation*}
C_{EK} = C_{EK}^{sat} S_w
\end{equation*}} & 
\parbox{7cm}{\begin{equation*}
-
\end{equation*}}
\\

\begin{minipage}{2.8cm}
\cite{darnet2004modelling}
\end{minipage}  & 
\parbox{4cm}{\begin{equation*}
C_{EK} = \frac{\epsilon \zeta}{\eta_w \sigma S_e}
\end{equation*}} & 
\parbox{7cm}{\begin{equation*}
-
\end{equation*}}
\\

\begin{minipage}{2.8cm}
\cite{revil2004streaming}$^{(2)}$
\end{minipage}  & 
\parbox{6cm}{\begin{equation*}
C_{EK} = C_{EK}^{sat} \frac{\beta_{(+)}(\sqrt{R^2+1}+R)+\beta_{(-)}(\sqrt{R^2+1}-R)}{\beta_{(+)}(\sqrt{R_S^2+1}+R_S)+\beta_{(-)}(\sqrt{R_S^2+1}-R_S)}
\end{equation*}} & 
\parbox{2cm}{\begin{equation*}
-
\end{equation*}}
\\

\begin{minipage}{3.2cm}
\cite{linde2007streaming} and \cite{revil2007electrokinetic}
\end{minipage}  & 
\parbox{6cm}{\begin{equation*}
C_{EK} = C_{EK}^{sat} \frac{ k_{rel}(S_w)}{S_w \sigma_{rel}(S_w)}
\end{equation*}} &
\parbox{5cm}{\begin{equation*}
\hat{Q}_v = \frac{\hat{Q}_v^{sat}}{S_w}
\end{equation*}}
\\

\begin{minipage}{3cm}
\cite{jackson2010multiphase}
\end{minipage}  & 
\parbox{4cm}{\begin{equation*}
C_{EK} = C_{EK}^{sat} \frac{S_e}{\sigma_{rel}(S_e)}
\end{equation*}} &
\parbox{7cm}{\begin{equation*}
\hat{Q}_v = \frac{\hat{Q}_v^{sat}S_e}{k_{rel}(S_e)}
\end{equation*}} 
\\

\begin{minipage}{3cm}
\cite{jougnot2012derivation}$^{(3)}$
\end{minipage}  & 
\parbox{4cm}{\begin{equation*}
C_{EK} = - \frac{\hat{Q}_v k}{\eta_w \sigma}
\end{equation*}} &
\parbox{7cm}{\begin{equation*}
\hat{Q}_v = \frac{ \int_{R_{min}}^{R_{S_w}} \hat{Q}_v^{R}(R,C_w^0) v^R(R)f_D(R)dR}{ \int_{R_{min}}^{R_{S_w}} v^R(R)f_D(R)dR}
\end{equation*}} 
\\

\begin{minipage}{3cm}
\cite{allegre20121}$^{(4)}$
\end{minipage}  & 
\parbox{4cm}{\begin{equation*}
C_{EK} = C_{EK}^{sat} S_e[1 + \beta(1-S_e)^\gamma]
\end{equation*}} &
\parbox{7cm}{\begin{equation*}
-
\end{equation*}}
\\

\begin{minipage}{3cm}
\cite{zhang2017streaming}$^{(5)}$
\end{minipage}  & 
\parbox{4cm}{\begin{equation*}
-
\end{equation*}} & 
\parbox{7cm}{\begin{equation*}
\hat{Q}_v = \hat{Q}_v^{sat}(p S_e^{-q} + r)
\end{equation*}}
\\
\noalign{\smallskip}\hline
\end{tabular}
\begin{minipage}{17.5cm}
$^{(1)}$ $C_v$ is coupling coefficient in the brine and $w$ the hydrodynamic resistance factor \\
$^{(2)}$ $R$ is the quotient between the excess charge in the pore water and the brine concentration; $\beta$, the mobility of the ions and $R_S$, the quotient between $R$ and $S_e$ \\
$^{(3)}$ $f_D$ is the capillary size distribution function \\
$^{(4)}$ $\beta$ and $\gamma$ are two fitting parameters \\
$^{(5)}$ $p$, $q$ and $r$ are the model fitting parameters
\end{minipage}
\end{table}

\subsection{Description of previous models}
\label{subsec:review}
\cite{wurmstich1994modeling} estimated SP signals from the coupling coefficient approach when two-phase (air-water) flow occurs. They predicted that $C_{EK}$ should increase with decreasing water saturation under the assumption that the nonwetting phase (air) is transported as bubbles. However, when both phases are homogeneously distributed throughout the pore space, this assumption is not valid.
\cite{perrier2000characterization} studied daily variations of electric potential data measured over several week which were interpreted as SP signal produced by unsaturated flow in the media. Based on the coupling coefficient approach, they inferred an empirical relationship to explain the dependence $C_{EK}(S_w)$ under the assumption that the electrical currents were affected by the unsaturated state similarly to the effect on the hydrological flow.
\cite{guichet2003streaming} established from measured SP signals in a sand column during a drainage experiment that the coupling coefficient decreases linearly with decreasing effective water saturation.
\cite{darnet2004modelling} performed synthetic 1D modeling of SP signals by estimating these signals from the coupling coefficient approach for air-water flow. They assumed that $C_{EK}$ increases when water saturation decreases by considering that the air is transported as bubbles.
\cite{revil2004streaming} measured the coupling coefficient as a function of water saturation of two consolidated rock samples. To describe the dependence of $C_{EK}(S_w)$, they developed a petrophysical relationship which includes the effect of surface electrical conductivity.

Based on a volume averaging approach, \cite{linde2007streaming} and \cite{revil2007electrokinetic} considered that as the effective water saturation decreases in the pore volume (while the amount of surface charge remains constant), the relative effective excess charge density should increase in the pore water. Then, they proposed that the effective excess charge density $\hat{Q}_v$ scales with the inverse of water saturation. Although this model has been shown to reproduce fairly well the behavior of the relative effective excess charge density in well-sorted media \citep[e.g.,][]{linde2007streaming,mboh2012coupled,jougnot2013self}, it does not seem appropriate for more heterogeneous media. As shown by \cite{jougnot2012derivation} and \cite{jougnot2015monitoring}, it seems to underestimate the increase of $\hat{Q}_v$ with decreasing saturation.

\cite{jackson2010multiphase} proposed a capillary tube model to study the electrokinetic coupling during a two-phase flow (water-air). Under the hypotheses of a thin EDL and that the charge on the surface of each capillary is constant, he derived expressions for the coupling coefficient and the effective excess charge density by considering the streaming currents under partial and total saturated conditions. The relative factor of these expressions $C_{EK}^{rel}$ and $\hat{Q}_v^{REV,rel}$ are expressed as functions of petrophysical properties. Note that the proposed model (Eq. \eqref{eq:Qv-proposedmodel}) is based on the effective excess charge dragged by the water flux and on the model to estimate $\hat{Q}_v$ in a single pore proposed by \cite{guarracinophysically}. Then, the $\hat{Q}_v^{REV,rel}$ expressions of both models are consistent although being derived from the two different approaches used to study the electrokinetic coupling. Nevertheless, the proposed model has a closed-form analytical expression which allows to estimate the total effective excess charge density from petrophysical parameters.

Following the works of \cite{jackson2008characterization, jackson2010multiphase} and \cite{linde2009comment} and taking into account the heterogeneous nature of soils, \cite{jougnot2012derivation} proposed a capillary-based approach to represent the porous media. In their model, the capillary size distribution can be inferred from the hydrodynamic curves of the porous medium: the water retention (WR) and relative permeability (RP) curve. The equivalent pore size distribution can be derived from either of the two curves. The effective excess charge density is calculated by numerically integrating the distribution of the pore water flux and the distribution of excess charge density within the capillaries (i.e., the flux averaging approach).

\cite{allegre20121} proposed an empirical relationship for the electrokinetic coupling coefficient inferred from drainage experiments performed within a column filled with a clean sand. This model predicts a non-monotonous behavior of the $C_{EK}$ coefficient as a function of water saturation. Note that the model of \cite{jougnot2012derivation} also predicts a non-monotonous behavior of the coupling coefficient as a function of the water saturation depending on the porous medium. It is the case for Fontainebleau sand \citep[i.e., the one used by][]{allegre20121} as it can be seen in  Fig. 6c and 6d of \cite{jougnot2012derivation}.

Recently, \cite{zhang2017streaming} proposed a power law function to model the relative effective excess charge density $\hat{Q}_v^{rel}$ as a function of effective saturation for two sandstone core samples during drainage and imbibition experiments. This empirical model depends on three parameters $p$, $q$ and $r$ which are determined by curve fitting to the mean values of $\hat{Q}_v^{rel}(S_e)$ at each value of water saturation for a given sample and displacement. In addition, they placed a constraint on two of the model parameters ($p+r=1$) since the relative effective excess charge density must equal 1 when the sample is fully saturated.

\subsection{Quantitative comparison with $\hat{Q}_v$ models}
\label{subsec:comparison-premodels}
In this section, we compare the proposed model with the effective excess charge models proposed by \cite{linde2007streaming}, \cite{jougnot2012derivation} and \cite{zhang2017streaming} described previously.

\begin{figure}
\includegraphics[width=1.1\textwidth,keepaspectratio=true]{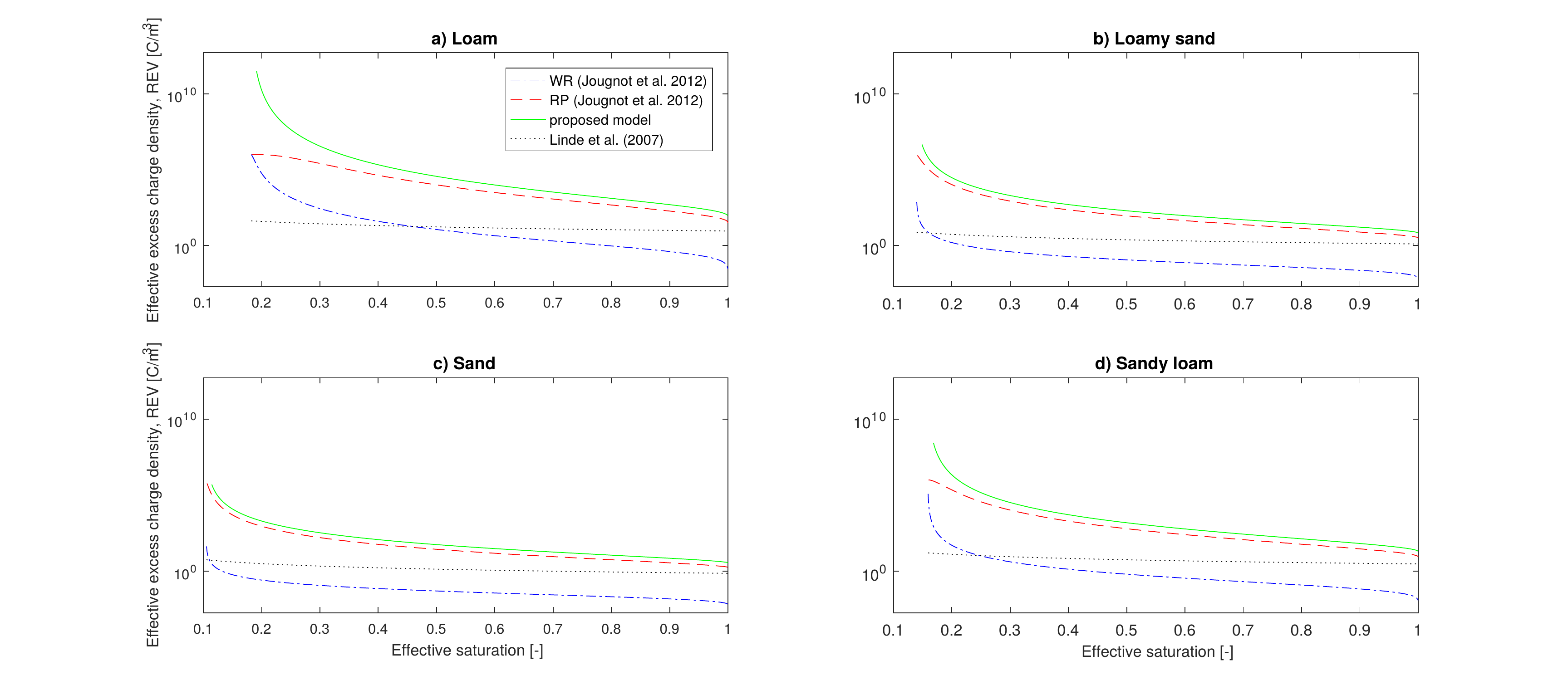}
\caption{Comparison between the proposed model (Eq. \eqref{eq:Qv-proposedmodel}), the water retention (WR) and relative permeability (RP) approaches from \cite{jougnot2012derivation} and the model of \cite{linde2007streaming} for different soil textures: a) loam, b) loamy sand, c) sand and d) sandy loam}
\label{fig:simulation}
\end{figure}

Firstly, Eq. \eqref{eq:Qv-proposedmodel} is compared with the models of \cite{linde2007streaming} and \cite{jougnot2012derivation}. Saturated effective excess charge values are required in order to estimate $\hat{Q}_v^{REV}$ using the $\hat{Q}_v^{REV,rel}$ model of \cite{linde2007streaming}. Then, the empirical relationship of \cite{jardani2007tomography} was considered to predict $\hat{Q}_v^{REV,sat}$. In their work, \cite{jougnot2012derivation} obtained $\hat{Q}_v^{REV}$ values for different soil textures from the WR and RP approaches using the hydrodynamic functions from \cite{van1980closed}. To ensure a consistent comparison, we also considered that constitutive model to estimate the permeability values in the proposed model. The parameters used for the different soil textures are listed in Table \ref{table:carselandparrish} \citep{carsel1988developing}. Then, we considered that the soils are saturated with a NaCl electrolyte at $T = 20^{\circ}C$ with a concentration of 5$\times 10^{-3}$ mol/L. Figure \ref{fig:simulation} shows the WR and RP models from the flux averaging of \cite{jougnot2012derivation}, the volume average approach from \cite{linde2007streaming} and the proposed model Eq. \eqref{eq:Qv-proposedmodel}. It can be seen that the proposed model shows an important increase of $\hat{Q}_v^{REV}$ values when water saturation decreases which is consistent with the previous models. However, the proposed model predicts much larger values for low saturations. The difference between the models is of several orders of magnitude (1 - 6 orders), being the strongest differences between the WR approach and the proposed model. The model of \cite{linde2007streaming} also differs significantly from the proposed model and it seems to underestimate the increase of $\hat{Q}_v^{REV}$ with decreasing saturation. Nonetheless, it can be noticed that the RP approach produces the smallest differences with the proposed model.

\begin{figure}
\centering
\includegraphics[width=0.8\textwidth,keepaspectratio=true]{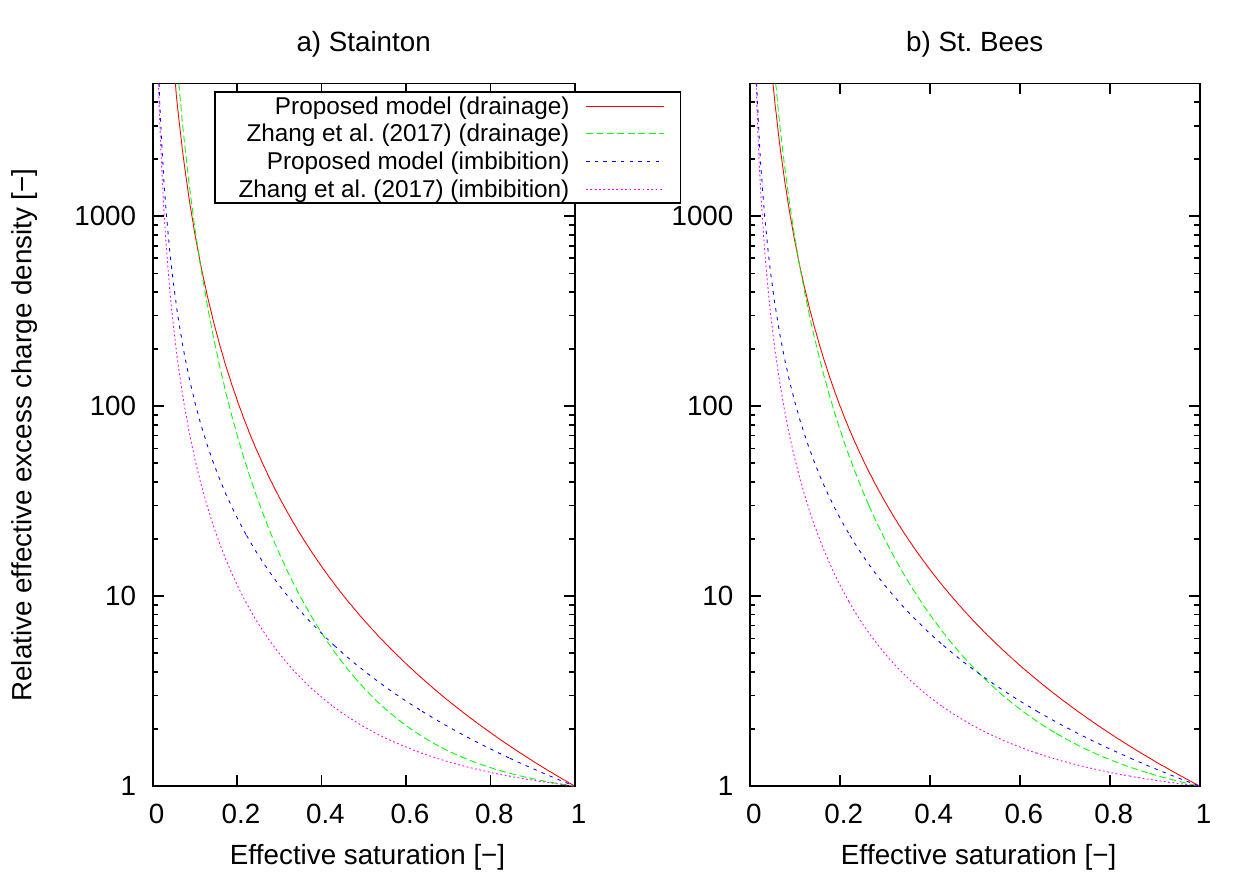}
\caption{Comparison between the relative effective excess charge density model (Eq. \eqref{eq:Qrel-kprop}) and the empirical model proposed by \cite{zhang2017streaming} for two sandstones: a) Stainton and b) St. Bees, during drainage and imbibition experiments}
\label{fig:zhang2017}
\end{figure}

Finally, the relative effective excess charge density model (Eq. \eqref{eq:Qrel-kprop}) is compared with the empirical model of \cite{zhang2017streaming}. Figure \ref{fig:zhang2017} shows $\hat{Q}_v^{REV,rel}$ as a function of effective saturation for two sandstones (Stainton and St. Bees) during drainage and imbibition experiments. The fitted parameters were $D=1.223$ and $\alpha = 4.995\times10^{-8}$ for drainage, and $D=1.01$ and $\alpha = 0.020$ for imbibition. We did not expect an exact fit of the models since they differ. However, both models are able to predict that the relative effective excess charge density strongly increases monotonically when water saturation decreases. 

\begin{table}
\centering
\caption{Typical hydrodynamic values for the van Genuchten-Mualem model as proposed by \cite{carsel1988developing}}
\label{table:carselandparrish}
\begin{tabular}[t]{lcccc}
\hline\noalign{\smallskip}
Texture & $\phi$ & $S_w^r$ & $m$ & $K (m/s)$  \\
\noalign{\smallskip}\hline\noalign{\smallskip}
Loam & 0.43 & 0.034 & 0.360 & 2.89$\times 10^{-6}$ \\
Loamy sand & 0.41 & 0.023 & 0.561 & 4.05$\times 10^{-5}$ \\
Sand & 0.43 & 0.019 & 0.627 & 8.25$\times 10^{-5}$ \\
Sandy loam & 0.41 & 0.027 & 0.471 & 1.23$\times 10^{-5}$ \\
\noalign{\smallskip}\hline
\end{tabular}
\end{table}

\section{Comparison with experimental data}
\label{sec:expdata}
In the present section, we test the proposed model against available experimental data from the research literature. These data sets consist of measured effective excess charge density-saturation or relative coupling coefficient-saturation values for different soil textures.

\subsection{Laboratory data}
\label{subsection:revil2004}
The proposed relative effective excess charge density model is tested with laboratory data obtained by \cite{revil2004streaming} and \cite{cherubini2018streaming}. The data from \cite{revil2004streaming} consist of two sets of relative coupling coefficient values for two dolomite core samples (named E3 and E39, diameter $\sim$3.8cm and length $<$ 8cm). It is worth  mentioning that the $\hat{Q}^{REV,rel}_v$ data values used to test our model were obtained from the $C^{rel}_{EK}$ measured values using the following relationship between them \citep{revil2004constitutive, revil2007electrokinetic}:

\begin{equation}
\hat{Q}^{REV,rel}_v = \frac{C^{rel}_{EK} \sigma^{rel}}{k_{rel}},
\label{eq:Q-Cek-rel}
\end{equation}
where the relative electrical conductivity $\sigma^{rel}$ was estimated using Archie's second law (\citeyear{archie1942electrical}) $\sigma^{rel} = \sigma(S_w)/\sigma^{sat} = S^n$, being $n$ the saturation index. The relative permeability $k_{rel}$ was calculated using Brooks and Corey model since \cite{revil2004streaming} fitted this model parameters to their experimental results. Table \ref{table:revil2004} lists the parameters used to estimate these electrical and hydraulic properties of the two samples \citep{revil2004streaming}. In order to provide a consistent comparison with the experimental data, $\hat{Q}_v^{REV,rel}$ values are estimated using Eq. \eqref{eq:Qrel-proposed}.

\begin{figure}
\includegraphics[width=1\textwidth,keepaspectratio=true]{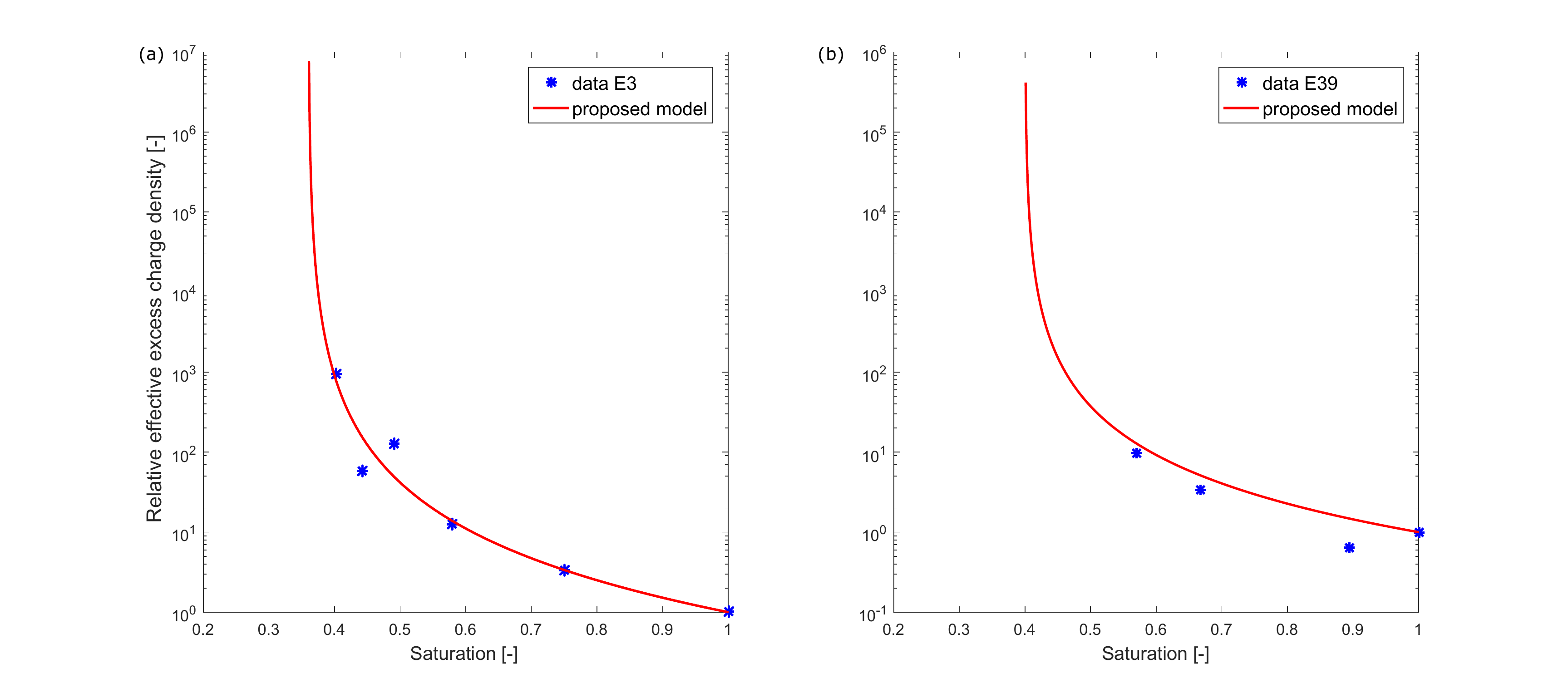}
\caption{Relative effective excess charge density as a function of saturation for data sets from two dolomites samples from \cite{revil2004streaming}}
\label{fig:revil-curves}
\end{figure}

Figure \ref{fig:revil-curves} shows the fit of Eq. \eqref{eq:Qrel-proposed} to predicted relative effective excess charge density values for the two dolomite samples. Parameter $D$ of the proposed model is fitted by an exhaustive search method and the best agreement between the predicted values and the model is obtained for $D=1.185$ ($RMSD = 0.567$) and $D=1.002$ ($RMSD = 0.482$), for E3 and E39 respectively. It can be seen that for both data sets, the proposed model fits fairly well for all the range of saturation values.

\begin{figure}
\includegraphics[width=1\textwidth,keepaspectratio=true]{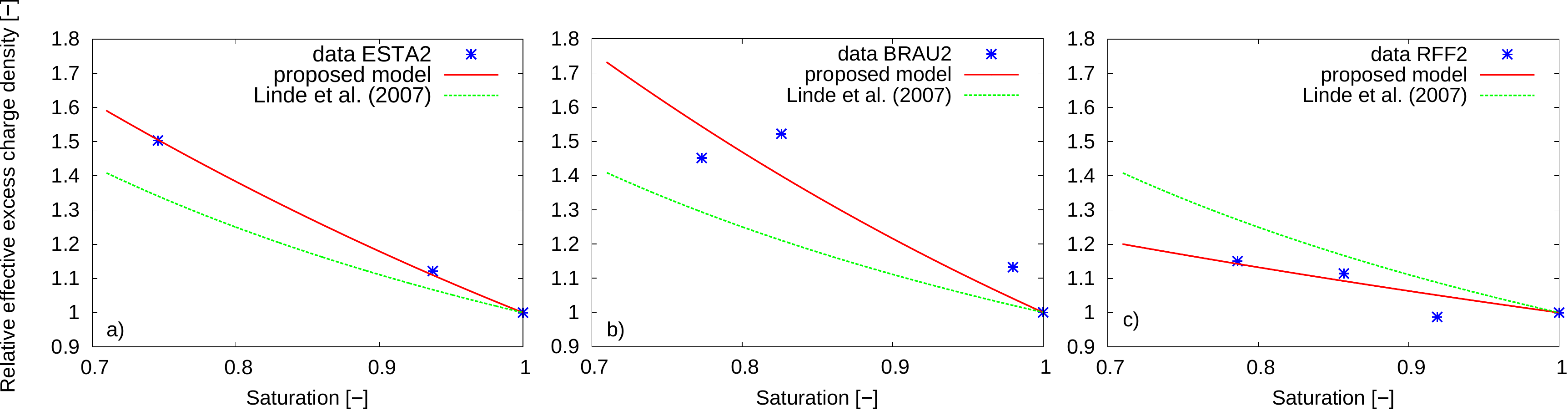}
\caption{Relative effective excess charge density as a function of saturation for data sets from three carbonate samples from \cite{cherubini2018streaming}.} 
\label{fig:cherubini}
\end{figure}

The data from \cite{cherubini2018streaming} consist on three sets of effective excess charge density values for three carbonate core samples (named ESTA2, BRAU2 and RFF2, diameter 3.9cm and length 8cm). In this case, the $\hat{Q}_v^{REV,rel}$ data values were obtained by normalizing the $\hat{Q}_v^{REV}$ data by the $\hat{Q}_v^{REV,sat}$ factor.
Figure \ref{fig:cherubini} shows the fit of Eq. \eqref{eq:Qrel-kprop} for the three carbonate samples and the model of \cite{linde2007streaming} is also shown. The best fitted parameters are listed in Table \ref{table:cherubini} which were found using the same method as for the previous laboratory data. It can be noted that Eq. \eqref{eq:Qrel-kprop} produces a fairly good agreement with the data sets for all the samples.

\begin{table}
\centering
\caption{Values of the hydraulic and electrical parameters used for the Brooks and Corey model (\citeyear{brooks1964hydraulic}) and for Archie's second law (\citeyear{archie1942electrical})}
\label{table:revil2004}
\begin{tabular}[t]{lccc}
\hline\noalign{\smallskip}
 & Electrical parameter & \multicolumn{2}{c}{Hydrological parameters} \\
Sample & n & $S_w^r$ & $\lambda$\\
\noalign{\smallskip}\hline\noalign{\smallskip}
E3 & 2.70 & 0.36 & 0.87\\
E39 & 3.48 & 0.40 & 1.65\\
\noalign{\smallskip}\hline
\end{tabular}
\end{table}

\begin{table}
\centering
\caption{Values of the fitted parameters for Eq. \eqref{eq:Qrel-kprop} for the data sets from \cite{cherubini2018streaming}}
\label{table:cherubini}
\begin{tabular}[t]{lcccccc}
\hline\noalign{\smallskip}
Sample & $S_w^{r \; *}$ & $\alpha$ & $D$ & RMSD \\
\noalign{\smallskip}\hline\noalign{\smallskip}
ESTA2 & 0.31 & 0.407 & 1.999 & 6.913$\times$10$^{-3}$ \\
BRAU2 & 0.30 & 0.351 & 1.999 & 8.965$\times$10$^{-2}$ \\
RFF2 & 0.35 & 0.684 & 1.010 & 3.387$\times$10$^{-2}$ \\
\noalign{\smallskip}\hline
\end{tabular}
\begin{minipage}{8.9cm}
$\qquad ^*$ Values taken from \cite{cherubini2018streaming}
\end{minipage}
\end{table}

\subsection{Field data}
\label{subsection:doussan2002}
We now test our model in a larger scale with experimental data acquired by \cite{doussan2002variations}. The experimental site consist of a lysimeter of about 9 m$^2$ surface area and 2 m depth packed with a sandy loam soil located in Avignon (France) in the experimental fields of the French National Institute for Agricultural Research (INRA). 
Following the analysis from \cite{jougnot2012derivation} on the lysimeter test from Doussan, we compare our model to the SP signals during 5 rainfall events. We converted the SP signals in $\hat{Q}_v^{REV}$ values using a 1D hypothesis and the process explained in \cite{jougnot2012derivation}.
As a result of the infiltration and evaporation of the rainwater, the water electrical conductivity $\sigma_w$ was changing with time between 0.06 S/m and 0.20 S/m. From this parameter data measured during the experiment and the use of the empirical formula of \cite{sen1992influence}, we obtained the ionic concentration $C_w^0$ of the free electrolyte that corresponds to each water electrical conductivity value (see Table \ref{table:doussan2002}). We then estimated the effective excess charge density using Eq. \eqref{eq:Qrel-proposed} to predict $\hat{Q}_v^{REV,rel}$ values. For $\hat{Q}_v^{REV,sat}$, we considered $\phi = 0.44$ \citep{doussan2002variations} and $\tau = 1.4$ as a representative value for the medium's tortuosity. This last parameter was estimated using the model of \cite{winsauer1952resistivity} to determine the hydraulic tortuosity as \citep[see the discussion in][]{guarracinophysically}:

\begin{equation}
\tau = \sqrt{F \phi}
\label{eq:tau_el}
\end{equation}
where $F$ is the formation factor, we considered $F = 4.54$ from \cite{doussan2002variations}. 

\begin{figure}
\includegraphics[width=1\textwidth,keepaspectratio=true]{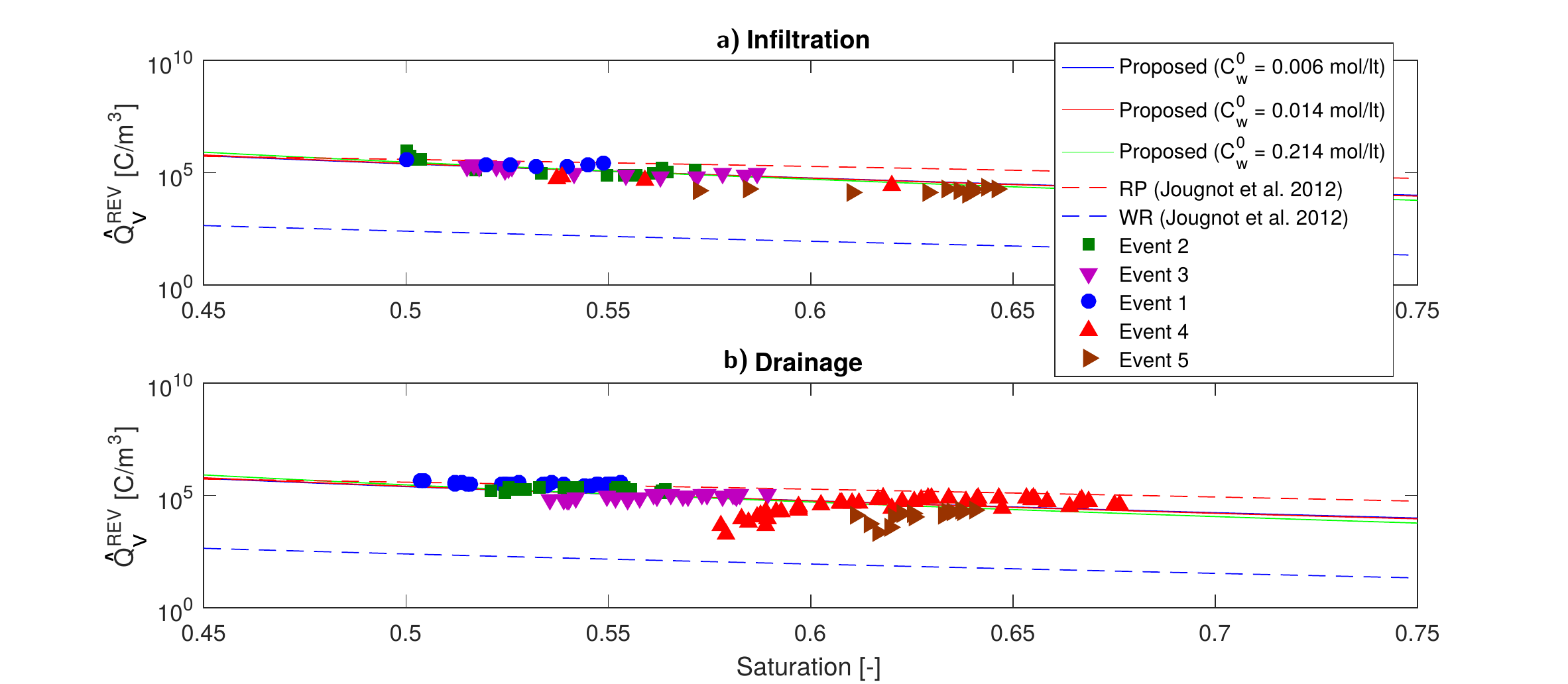}
\caption{Effective excess charge density as a function of saturation during the five rainfall events considering infiltration and drainage phases. The solid lines represent the proposed model for different values of ionic concentration $C_w^0$. The dashed lines represents the WR and RP approaches of \cite{jougnot2012derivation}}
\label{fig:doussan-curve}
\end{figure}

Figure \ref{fig:doussan-curve} shows the different rainfall events fitted with the proposed model for three different $C_w^0$ values corresponding to the maximum, minimum and medium values of $\sigma_w$. Table \ref{table:doussan2002} lists the resulting parameter $D$ for each $C_w^0$. Figure \ref{fig:doussan-curve} also shows the approaches of \cite{jougnot2012derivation} and it can be observed that the proposed model fits well all the events and provides much more better results than the WR approach. Moreover, it also produces better predictions of $\hat{Q}^{REV}_v$ values than the RP approach. 

\begin{table}
\centering
\caption{Values of the fitted parameter $D$ for the different values of water conductivity}
\label{table:doussan2002}
\begin{tabular}[t]{cccc}
\hline\noalign{\smallskip}
$\sigma_w$ (S/m) & $C_w^0$ (mol/L) & $D$ & $RMSD$ \\
\noalign{\smallskip}\hline\noalign{\smallskip}
0.06  & 0.006 & 1.749 & 3.219 \\
0.13  & 0.014 & 1.758 & 3.188 \\
0.20  & 0.214 & 1.793 & 3.056 \\
\noalign{\smallskip}\hline
\end{tabular}
\end{table}

\section{Discussion and conclusions}
\label{sec:discuss}
A physically based model for estimating the effective excess charge density under partially saturated conditions has been developed in the present study. The model was derived assuming that the porous media is represented by a bundle of tortuous capillary tubes with a fractal pore-size distribution and filled by a binary symmetric 1:1 electrolyte. Hydraulic and electrokinetic properties of the porous media were first estimated for a single pore at microscale and then upscaled to a REV assuming that a pore can be fully water saturated or completely dry. The resulting expression of $\hat{Q}_v^{REV}$ is a function of water effective saturation and relative permeability, and has an analytical closed-form that depends on chemical parameters of the pore water ($C_w^0$, $l_D$, $\zeta$) and hydraulic parameters ($\tau$,$\phi$,$k$). The proposed model is the extension of the model derived by \cite{guarracinophysically} for saturated conditions. Therefore, our results show that estimates of the effective excess charge density can be easily extrapolated from saturated porous media to partially saturated conditions by introducing a $\hat{Q}_v^{REV,rel}$ factor which depends on the effective saturation and relative permeability of the media. It should be noted that the resulting expression of $\hat{Q}_v^{REV,rel}$ (Eq. \eqref{eq:Qrel-generic}) is found consistent with the expression proposed by \cite{jackson2010multiphase} based on the framework of the coupling coefficient approach. In addition, the proposed model provides a relationship between $\hat{Q}_v^{REV}$ and $k k_{rel}$ (Eq. \eqref{eq:jardani-unsat}) which is similar to the empirical relationship of \cite{jardani2007tomography}. This effective excess charge density-permeability relationship extend to partially saturated conditions the recent analytical expression derived by \cite{guarracinophysically} under saturated conditions.

The proposed relative effective excess charge density model depends only on the hydraulic properties of the media. The analysis on different soil textures showed significant differences in the $\hat{Q}_v^{REV,rel}$ values (Fig. \ref{fig:textures}). Indeed, it could be observed that for a given value of effective saturation, the maximum values of $\hat{Q}_v^{REV,rel}$ are associated to fined soil textures. The performance of $\hat{Q}_v^{REV,rel}$ was also tested when different constitutive models are used to predict relative permeability. The comparison between van Genuchten, Brooks and Corey and the proposed relative permeability (Eq. \eqref{eq:kdeSyRmin-max}) models showed that all of them can provide good estimates of $\hat{Q}_v^{REV,rel}$ values (Fig. \ref{fig:testkrel}).

The proposed model represents an improvement to predict $\hat{Q}_v^{REV}$ values over the previous models of \cite{linde2007streaming} and \cite{jougnot2012derivation} also derived from the effective excess charge approach. When compared to the model of \cite{jougnot2012derivation}, the proposed model produces a better match to the relative permeability approach rather than to the water retention (Fig. \ref{fig:simulation}). The comparison of the proposed $\hat{Q}_v^{REV,rel}$ model against the empirical relationship proposed by \cite{zhang2017streaming} showed that both models predict the behavior of increasing $\hat{Q}_v^{REV,rel}$ when effective saturation decreases, although we did not expect a perfect match since the strong differences between the models. 

The proposed $\hat{Q}_v^{REV}$ and $\hat{Q}_v^{REV,rel}$ models were validated using experimental laboratory and field data, respectively. In both cases, the models were able to satisfactorily predict the magnitude of the data (Fig. \ref{fig:revil-curves}, \ref{fig:cherubini} and \ref{fig:doussan-curve}). Indeed, in the case of the field data, the $\hat{Q}_v^{REV}$ model showed better agreement than the water retention and relative permeability approaches from \cite{jougnot2012derivation}.

This study allowed the development of a new model to describe SP signals from the framework of the effective excess charge. This relationship is a explicit function of effective saturation while also depending on petrophysical parameters of the porous media. The analytical closed-form expression of this new model is easy to evaluate and its numerical implementation is straightforward. These observations may have practical implications for opening up to the use of the SP method to study physical phenomena occurring in the unsaturated zone such as infiltration \citep[e.g.,][]{jougnot2015monitoring}, evaporation, CO$_2$ flooding \citep[e.g.,][]{busing2017numerical}, or even contaminant fluxes \citep[e.g.,][]{linde2007inverting}.

\newpage


\end{document}